\documentclass[%
pra, 
 mph,%
 amsmath,amssymb,
 reprint,%
]{revtex4-2}

\usepackage{graphicx}
\usepackage{dcolumn}
\usepackage{bm}

\usepackage{graphicx}
\usepackage{dcolumn}
\usepackage{bm}
\usepackage{hyperref}
\usepackage[caption=false]{subfig}

\usepackage[mathlines]{lineno}
\modulolinenumbers[5]
\linenumbers\relax 

\begin{document}
\nolinenumbers
\preprint{AAPM/123-QED}

\title{Multi-MeV electrons from above-threshold ionization of the neon K-shell}

\author{A. Yandow}
\affiliation{Center for High Energy Density Science, The University of Texas at Austin, 2515 Speeday Stop C1600, Austin, TX 78712}%
\affiliation{Lawrence Livermore National Laboratory, Livermore, CA 94551}
\email{yandow1@llnl.gov}
\author{T. N. Ha}%
\author{C. Aniculaesei}
\author{H. L. Smith}
\author{C. G. Richmond}
\author{M. M. Spinks}
\author{H. J. Quevedo}
\author{S. Bruce}
\author{M. Darilek}
\author{C. Chang}
\author{D. A. Garcia}
\author{E. Gaul}
\author{M. E. Donovan}
\author{B. M. Hegelich}
\author{T. Ditmire}
\affiliation{Center for High Energy Density Science, The University of Texas at Austin, 2515 Speedway Stop C1600, Austin, TX 78712}%

\date{\today}
\begin{abstract}
We present measurements of integrated electron energies produced by above-threshold ionization (ATI) of neon in a laser field with intensity exceeding $10^{20}$ W/cm$^{2}$. We observe electrons with energy exceeding 10 MeV ejected in the laser forward direction above a threshold intensity of 2 $\times 10^{20}$ W/cm$^{2}$. We compare to  ATI models using both tunneling (ADK-PPT) and barrier suppression ionization and observe the onset of ATI at a higher threshold intensity than predicted by these models. 
\end{abstract}

\maketitle
        Above-threshold ionization (ATI) is a fundamental response of an atomic system to an intense flux of photons. The first experimental evidence of ATI showed the absorption of seven photons in a six-photon multiphoton ionization pathway of xenon \cite{Agostini1979a}. As near-infrared laser intensity increases beyond $10^{14}$ W/cm$^{2}$, ATI can be well-described by a quasi-classical two-step model. The ionization process can be described by the Ammosov-Krainov-Delone and Perelomov-Popov-Terent'ev (ADK-PPT) model \cite{Ammosov1986}\cite{Perelomov1966}, where the photons sum coherently to create a strong electric field which liberates electrons through a quasi-static tunneling process. The ATI electron is then ``born'' into the laser field with initial conditions consistent with the ADK-PPT tunneling model, and the absorbed laser energy can be found by integrating the classical Lorentz force equations. The ADK-PPT ionization rate model has been well-validated by measurements of ion charge states produced in the laser focus at nonrelativistic intensities \cite{Augst1989a}. Extension of these experiments to intensity above $10^{20}$ W/cm$^{2}$ has been elusive even with advances in laser technology due to the limited repetition rate of high-energy ultrafast lasers and the gas density limits imposed by conventional time-of-flight experiment designs \cite{Link2006}.

        Yamakawa \textit{et al.} and Chowdhury \textit{et al.} performed precision measurements of highly-charged noble gas ions at relativistic laser intensities exceeding $2 \times 10^{19}$ W/cm$^{2}$\cite{Chowdhury2001}\cite{Yamakawa2003}. Both authors exploited the strongly nonlinear dependence of tunneling ionization probability on laser intensity to calculate a model intensity from the relative charge state yields. The intensity computed by Chowdhury \textit{et al.}  was at the estimated lower bound of their experimental intensity. Yamakawa \textit{et al.}  found unexpectedly that the laser intensity calculated from the ADK-PPT tunneling model depended on the noble gas ion species when the laser parameters were held constant, with the intensity calculated from the model decreasing systematically with increasing atomic number. Recent modeling suggests the ionization of helium-like ions is more robust to uncertainties in the ionization modeling \cite{Ciappina2020a}\cite{Lotstedt2020a}, motivating investigation into ionization of the neon K-shell.

        In this Letter we explore the observation of K-shell neon charge states produced in a laser focus by detecting the high-energy ATI electrons produced by the laser-ion interaction. We select high-energy ATI electrons as our experimental observable because direct laser acceleration of the highly-charged ions will severely degrade the resolution of time-of-flight spectrometers as intensity approaches $10^{21}$ W/cm$^{2}$\cite{Yandow2019}. The behavior of relativistic ATI electrons is well-characterized by experiment \cite{Dichiara2008}\cite{Ekanayake2013}\cite{Moore1995}, with higher-energy electrons confined to a smaller forward cone in the laser forward direction at an angle
\begin{equation}
tan(\theta) = \sqrt{\frac{2}{\gamma - 1}}
\end{equation}
from the laser propagation direction\cite{Moore1995}. Modulation of the ATI electron energy spectrum and spatial distributions induced by the gaps in appearance intensity between different atomic shells has also been confirmed experimentally in argon and xenon \cite{Ekanayake2013}. The large gap in ionization potential between the L-shell ($<$ 239 eV) and K-shell electrons (1196 eV and 1362 eV) of neon result in the K-shell electrons being ``born'' into a field nearly two orders of magnitude higher in intensity, and the K-shell ATI electrons that interact with the laser field at peak strength will be ejected in a narrower cone centered on the laser forward direction than the L-shell electrons that will be ponderomotively expelled by the leading edge of the laser pulse and will therefore attain lower energies.

\begin{figure}[t!]
\includegraphics[width = \linewidth]{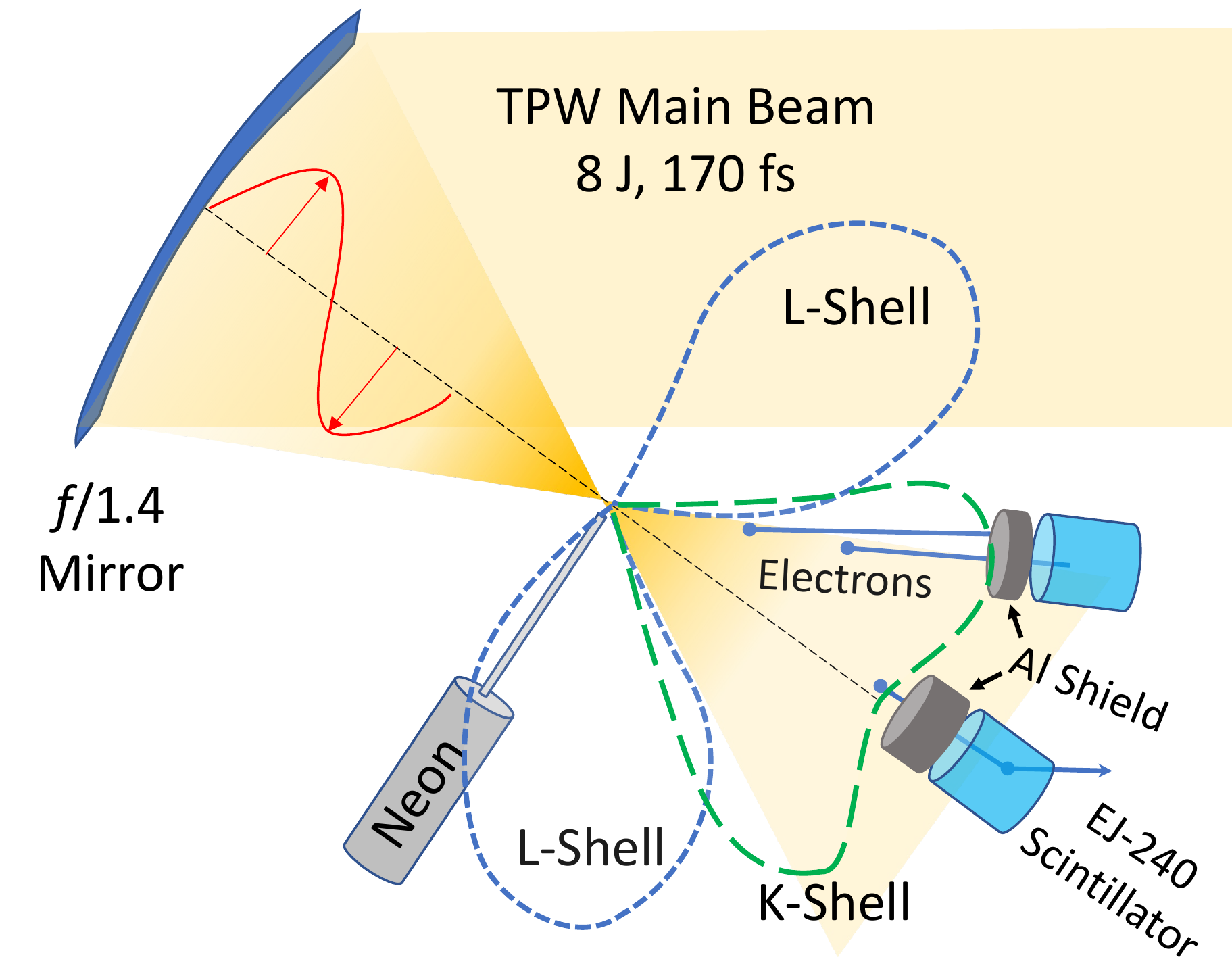}
\caption{\label{diagram} A conceptual diagram of the experimental setup. Two detectors discussed in this Letter are oriented at angles of $0^{\circ}$ and $30^{\circ}$ from the laser forward direction and in the laser polarization plane. The dashed lobes demonstrate angular separation of the L-shell and K-shell electrons. This drawing is not to scale. Color figures available online.}
\end{figure}

        Figure \ref{diagram} shows a simplified diagram of our experimental setup, which is described in greater detail elsewhere \cite{YandowPRA2023}. A multi-Joule, linearly polarized, ultrafast laser pulse produced by the rod amplifier of the Texas Petawatt laser is focused with an $f$/1.4 off-axis parabolic mirror, focusing to a spot with central maximum full-width at half-maximum measured to be $2.6 \pm 0.2$ $\mu$m with optimal wavefront. The intensity at the focal plane in the target chamber is estimated using indirect measurements of the wavefront, an estimated pulse duration deconvolved from a second-order autocorrelation measurement assuming a Gaussian pulse shape, and a calibrated energy measurement. A peak intensity of $5 \times 10^{20}$ W/cm$^{2}$ is attained in this configuration. Intensity was scanned by decreasing the laser energy with the laser wavefront and pulse duration remaining optimized.

        The focused laser pulse interacts with a low-density plume of neon gas near the target chamber center that is introduced by a flow-calibrated orifice with a diameter of 100 $\mu$m backed with 60 torr of ultra high-purity neon gas. We estimate the maximum density of the gas in the interaction volume to be about $3 \times 10^{14}$ cm$^{-3}$ using Ansys Fluent\cite{AnsysF} simulation of the steady-state gas flow into vacuum, below the threshold density for collective plasma effects. The electrons produced by the laser-ion interactions are detected by scintillating calorimeter detectors placed around the laser focus. The three detectors discussed in this Letter were oriented along the polarization plane at 30$^{\circ}$ from the laser forward direction, along the laser forward direction, and at a control position 110$^{\circ}$ from the laser forward direction and out of the polarization plane.
        
\begin{figure}[t!]
\begin{minipage}{\linewidth}
\subfloat{\label{main:fig2a}\includegraphics[width = \linewidth]{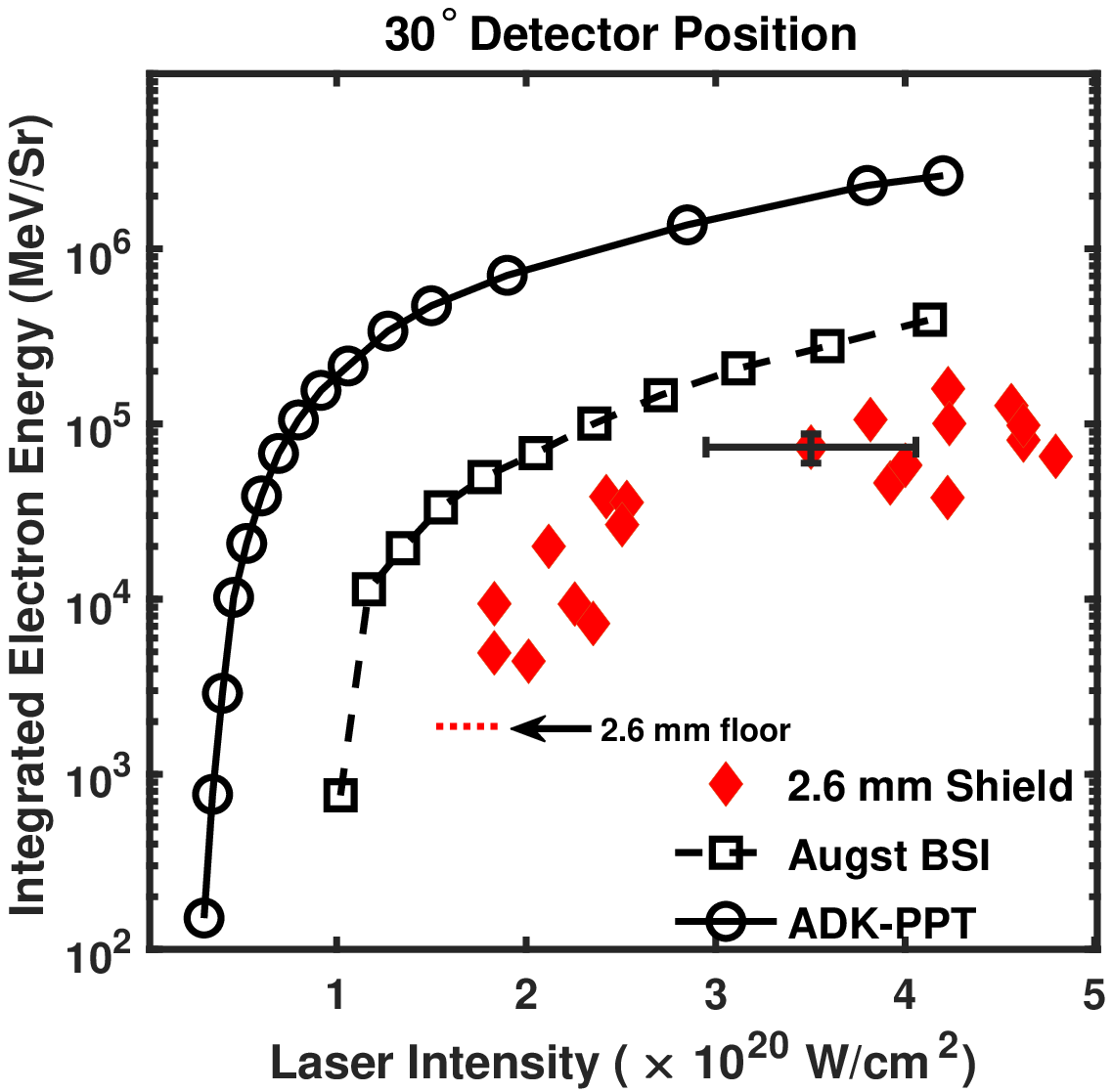}}

\subfloat{\label{main:fig2b}\includegraphics[width = \linewidth]{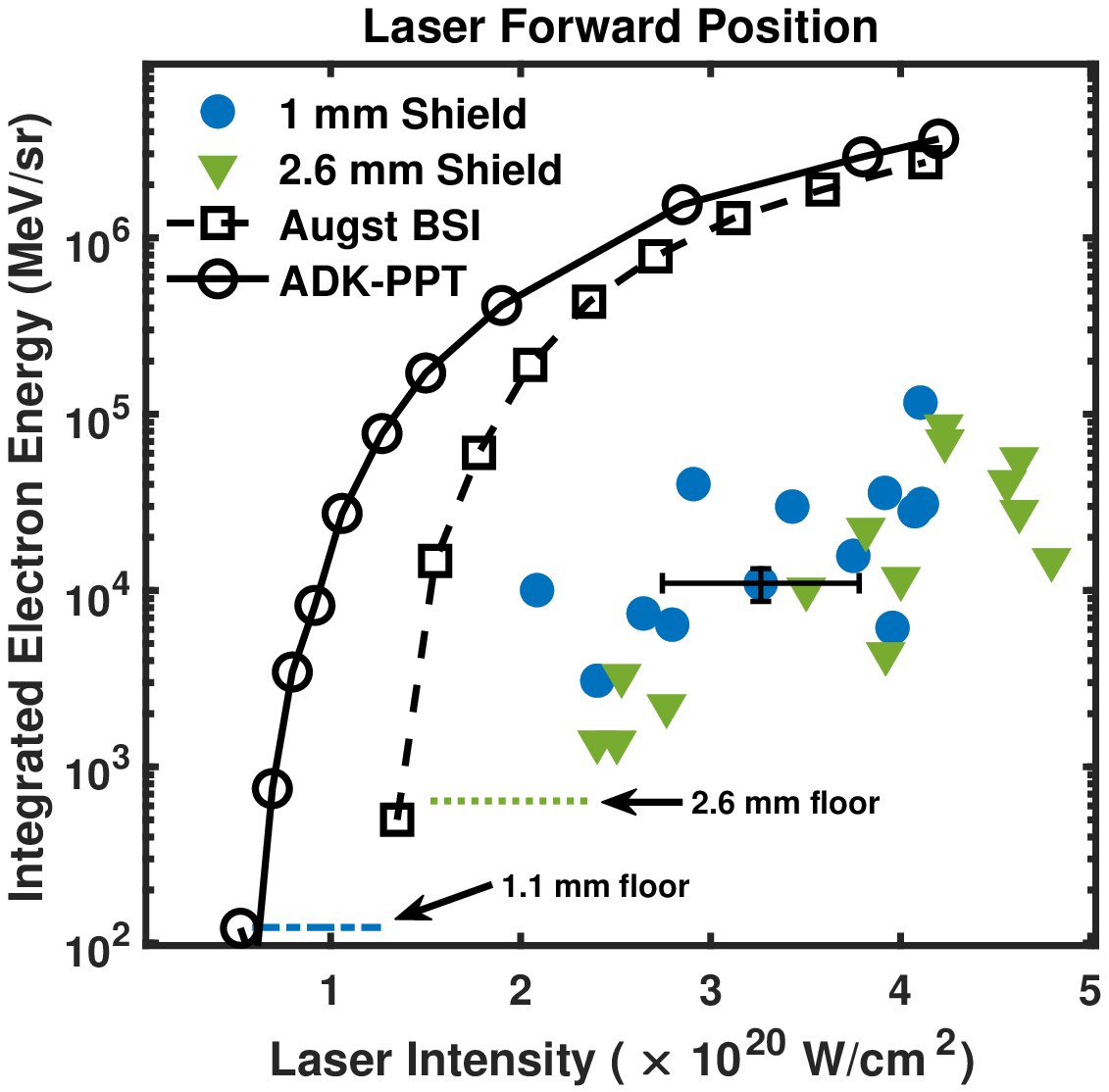}}
\end{minipage}
\caption{\label{fig:fig2} Electron energy deposited in the scintillator at the (a)$30^{\circ}$ position and (b) $0^{\circ}$ position (b). Shield thicknesses of 1 mm and 2.6 mm have $50\%$ efficiency cutoffs at 1.4 MeV and 2.8 MeV, respectively. Predictions of the ADK-PPT model (solid curve) and Augst BSI model (dashed curve) are presented alongside, with simulation intensities at the marker points. Color figures available online. The floors mark the maximum values and intensity ranges for measurements falling below the lowest detector charge threshold in each shielding configuration.} 
\end{figure}
        
         Each detector consisted of a 50 mm diameter, 40 mm long cylinder of long-lifetime (285 ns) scintillating plastic (Eljen Technologies EJ240) coupled to a photomultiplier tube with a tapered voltage divider for optimal pulse linearity. The scintillator plastic and photomultipier tubes (PMT) were encased in a vacuum-compatible PTFE housing that was made light-tight with colloidal graphite and aluminum foil. The relatively large solid angle ($\sim$ 0.03 steradians) subtended by the detectors  captured several hundred ATI electrons at each detector, enabling accurate calorimeter energy measurements with only a few shots at each laser intensity. The output current pulse from each photomultiplier tube was recorded on a Tektronix TDS5054 oscilloscope and digitally filtered to eliminate ringing from on-shot electromagnetic noise. The upper and lower charge thresholds at each voltage were estimated from the breakdown of the linear relationship between current amplitude and integrated charge due to pulse saturation and residual ringing, respectively.

        We performed a series of control experiments to confirm our signal originated from K-shell ATI electrons. We verified the signal was not due to electromagnetic pulse effects by verifying it disappeared when the gas flow was turned off. We verified the signal from forward-directed radiation was at least 500 times greater than backward-detected radiation at the control detector position. We swapped the detectors located at the $30^{\circ}$ and control positions to confirm the observed signal was due to radiation detected in the laser forward direction and not a detector-specific artifact. A series of control shots was also taken using helium gas as a target to simulate the laser interaction with the L-shell electrons. We observed no repeatable signal that would correspond to L-shell electrons with energy greater than 1.4 MeV at the 0$^{\circ}$ position and 2.8 MeV at the 30$^{\circ}$ position.

        Figures \ref{main:fig2a} and \ref{main:fig2b} show our integrated ATI electron energy yields at the $30^{\circ}$ and $0^{\circ}$ detector positions. Aluminum filters with thicknesses of 1 mm and 2.6 mm are inserted in front of the detectors to stop electrons with energy below 1.4 MeV and 2.8 MeV, respectively, to eliminate low-energy L-shell electrons observed scattered toward the laser forward direction \cite{YandowPRA2023}. We observe a similar intensity threshold effect at both positions, where the energy carried forward by ATI electrons becomes measurable above $2 \times 10^{20}$ W/cm$^{2}$ and increases rapidly with intensity. The minimum signal level, below which detector ringing renders our estimated uncertainty unreliable, is marked for both shielding configurations presented in Figure \ref{main:fig2b}, showing that the ATI electron signal drops off more than an order of magnitude at the K-shell ionization threshold intensity. A scaling transition characteristic of tunneling ionization \cite{Augst1989a}\cite{Augst1991} is visible at an intensity around $3 \times 10^{20}$ W/cm$^{2}$ in Figure \ref{main:fig2a}, where the measured ATI electron energy yield transitions to a power-law intensity dependence dominated by the volume of the focal region exceeding the ionization threshold intensity. Below this intensity, the integrated ATI electron energy dependence is dominated by the probability of K-shell ionization in the most intense region of the laser focus, which is a highly nonlinear function of intensity.  
        
        We observe the K-shell ionization threshold intensity is nearly double the barrier suppression intensity, and that both the ADK-PPT and Augst barrier suppression models described in detail elsewhere \cite{YandowPRA2023} underestimate the K-shell ionization intensity threshold. The primary sources of error originate from the detector energy calibration ($\sim 20\%$) and our method of calculating intensity ($\sim 15-20 \%$), and are not sufficient to explain the discrepancy between the observed ATI electron energy yields and the Monte Carlo models.
        
\begin{figure}[t!]
\begin{minipage}{\linewidth}
\subfloat{\label{main:fig3a}\includegraphics[width = \linewidth]{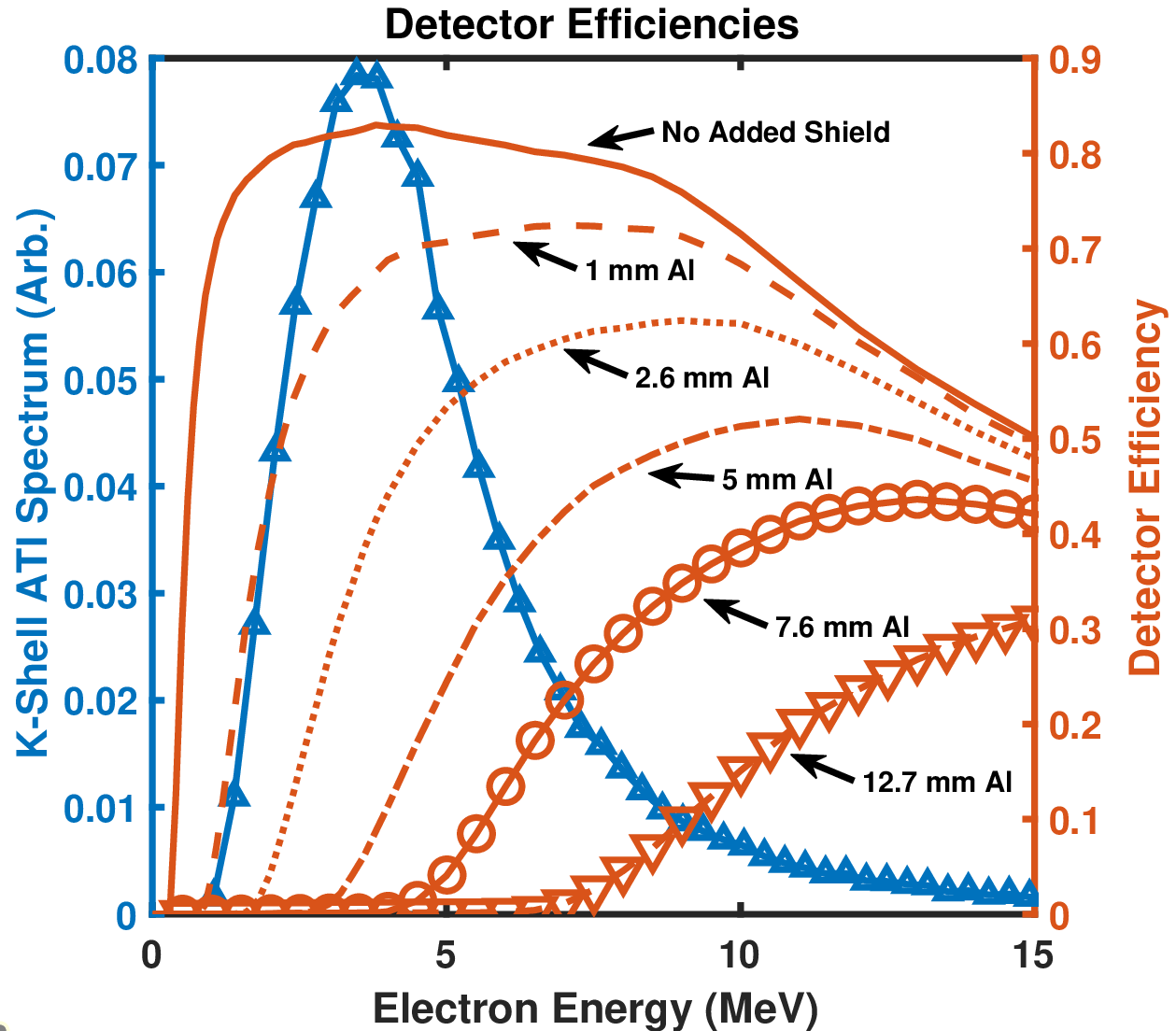}}

\subfloat{\label{main:fig3b}\includegraphics[width = \linewidth]{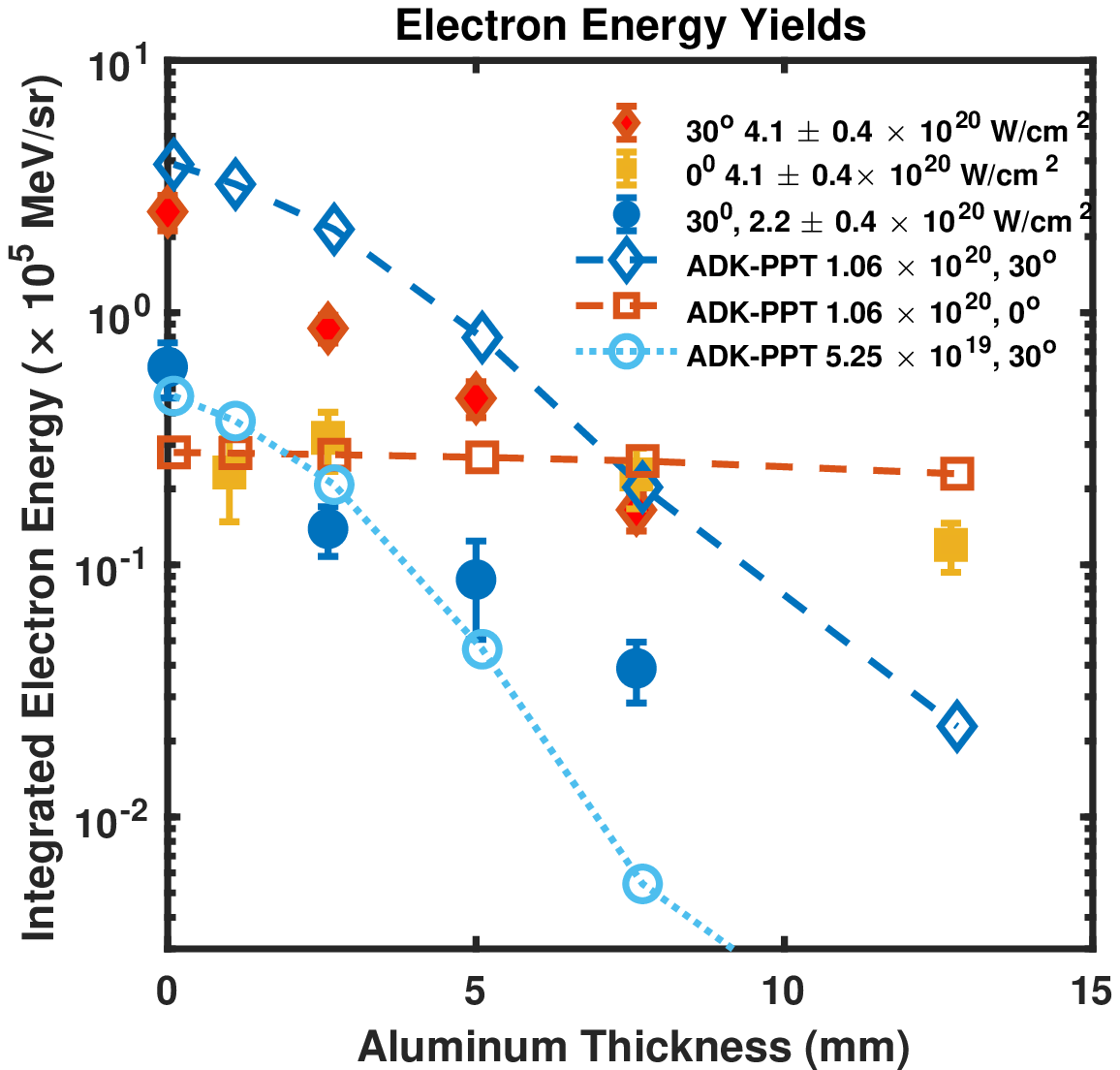}}
\end{minipage}
\caption{\label{fig:epsart} a) Detector efficiencies at different aluminum thicknesses (right axis)  and a simulated K-shell ATI electron spectrum using the ADK-PPT model at intensity of $1.06 \times 10^{20}$ W/cm$^{2}$ (left axis) shown for comparison. b) Integrated ATI electron energy at the two positions at two different average intensities, with ADK-PPT simulation predictions (open markers) for comparison.}
\end{figure}

        The limited number of laser shots and the low density of target gas prevented measurement of the electron spectrum using a magnetic spectrometer. We placed a series of aluminum filters of different thicknesses in front of the scintillating plastic and took repeated measurements to gain information about the energy spectrum and the maximum electron energy. Figure \ref{main:fig3a} shows a series of detector efficiency curves calculated using G4beamline \cite{Roberts2007} for different aluminum filter thicknesses, with thicker filters shielding the scintillating plastic from lower-energy electrons. Although the lack of a sharp cutoff in the energy efficiency curves makes it impossible to invert our integrated measurements to obtain a unique electron energy spectrum, we can gain qualitative information on the shape of the spectrum and estimate a maximum ATI electron energy range by comparing the ratio of the measured integrated electron energy for the two thickest shields and to the ratio of the respective efficiency curves.  
        
\begin{figure}[b!]
\includegraphics[width = \linewidth]{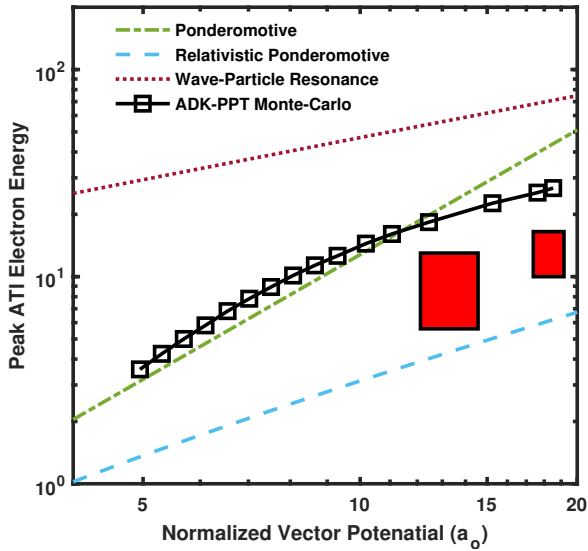}
\caption{\label{fig4} Comparison between the peak ATI electron energies predicted by different models and our experimental results. Curves are analytic models (see Refs. \cite{Gordon2017} and \cite{Maltsev2003a}) and the markers represent the average of the top 10$\%$ most energetic electrons in the ADK-PPT Monte-Carlo simulations. The shaded red regions represent peak ATI electron consistent with the data in Fig. \ref{main:fig3b}}
\end{figure}
        
        Figure \ref{main:fig3b} shows the measured energy yields at the $30^{\circ}$ and on-axis detectors, where the modeling predicts the highest number of K-shell ATI electrons. At an average intensity of $4.1 \pm 0.4 \times 10^{20}$ W/cm$^{2}$, we found the most favorable comparison was to the ADK-PPT ATI electron model with a model intensity of $1.06 \times 10^{20}$ W/cm$^{2}$. The Augst BSI model predicted a significantly higher ratio of energy yield in the laser forward direction than the ADK-PPT model at every intensity, and a model intensity range consistent with the energy yields observed at both detector positions could not be found for the Augst BSI model \cite{YandowPRA2023}. From the experimental measurements at $30^{\circ}$ at two intensities ($(4.1 \pm 0.4) \times 10^{20}$ W/cm$^{2}$ and $(2.2 \pm 0.4) \times 10^{20}$ W/cm$^{2}$), the ADK-PPT model falls off quicker with increasing shield thickness than the measured yields because the model 
significantly underestimates the proportion of electrons with energy $>$ 6 MeV at both model intensities. We cannot make a conclusive observation about the K-shell electrons with energy $ < 3$ MeV because our helium control shots indicate that lower-energy L-shell electrons are scattered further into the laser forward direction than predicted by modeling \cite{YandowPRA2023}. 
        
        The maximum ATI electron energy ranges consistent with our measured integrated energies at these two intensities are 5.6-13 MeV and 10-16 MeV, respectively. Figure \ref{fig4} compares these ranges to different analytic models of peak ATI electron energy: the ponderomotive, relativistic ponderomotive, and superponderomotive ``wave-particle resonance'' \cite{Gordon2017}\cite{Maltsev2003a} models. The experimentally determined ranges fall between the relativistic and nonrelativistic ponderomotive models, with the Monte Carlo model overestimating the maximum ATI electron energy by a factor or 2-3. The Monte-Carlo modeling likely overestimates the integrated ATI electron energies because a Gaussian laser focus is assumed to make the model computationally tractable. The predicted electron energy at the $0^{\circ}$ position will be overestimated because higher-order spatial modes will dephase an electron co-propagating with the laser field more quickly due to stronger Gouy phase shift associated with higher-order modes, decreasing the maximum electron energy along the laser direction\cite{Maltsev2003a}. Our experimental finding that the maximum ATI electron energy falls between the ponderomotive and relativistic ponderomotive models raises an important theoretical question about whether the superponderomotive scaling of the maximum ATI electron energy at the onset of ``wave-particle resonance'' predicted by D. F. Gordon \textit{et al.}  \cite{Gordon2017} and demonstrated by our Monte-Carlo modeling in Figure \ref{fig4} would be a feature of a non-Gaussian laser focus. 
        
        We find that neither the ADK-PPT Monte-Carlo model nor the Augst BSI model provide a fully consistent quantitative description of integrated ATI electron energy measurements. While a significant threshold intensity shift is unexpected for the ionization of a simple helium-like atom, it is not inconsistent with the other ionization experiments at relativistic intensities. Yamakawa \textit{et al.}  compared direct measurements of ion charge states to the ADK-PPT model and found that the ADK-PPT model predicted a laser intensity that was a factor of 2-8 lower than the indirectly calculated laser field intensity of $2.6 \times 10^{19}$ W/cm$^{2}$, with heavier noble gas ions consistent with lower field intensities \cite{Yamakawa2003}. While laser intensity has been claimed to exceed the barrier suppression intensity of the neon K-shell states since at least 2006 and charge states as high as Kr$^{24+}$ have been observed \cite{Akahane2006a}, our work represents the first inference of neon K-shell ionization that has been reported in the literature to our knowledge and the first detection of ATI electrons exceeding 10 MeV, corresponding to the absorption of $10^{7}$ photons during the ionization process. 
        
        The detection of ATI electrons from K-shell states of argon ($\sim 3 \times 10^{21}$ W/cm$^{2}$) and krypton ($\sim 10^{23}$) W/cm$^{2}$ are relatively straightforward extensions of our work made possible by development of repetition-rated multi-petawatt laser systems \cite{Rus2015}. The interaction of Petawatt laser pulses with highly-charged ions can generate ATI electrons with energy exceeding 100 MeV\cite{Yandow2019}\cite{Maltsev2003a}, would exit the focus in a small forward cone that could be monitored with a magnetic spectrometer or a scintillating detector array. Extension of this basic experimental design and the development of computational techniques to model the interaction of ultra-relativistic electrons with realistic spatio-temporal laser fields are active areas of research necessary to further develop a direct intensity measurement technique using ATI electrons from highly-charged ions in an intense laser field. 
        
        A. Y. acknowledges helpful conversations with E. Chowdhury regarding the design of this experiment. This work was supported by the DOE, Office of Science, Fusion Energy Sciences under Contract No. DE-SC0021125: LaserNetUS: A Proposal to Advance North America's First High Intensity Laser Research Network, the Air Force Office of Scientific Research through Awards No. FA9550-14-1-0045 and No. FA9550-17-1-0264, the National Nuclear Security Agency (NNSA) through Award No. NA0002008, and was prepared by LLNL under Contract DE-AC52-07NA27344.  A. Y. gratefully acknowledges the generous support of the Jane and Michael Downer Fellowship in Laser Physics in Memory of Glenn Bryan Focht.

\bibliography{prl_bib_vFINAL.bib}  
\end{document}